\documentclass[aps, superscriptaddress, reprint, longbibliography]{revtex4-1}
\usepackage{graphicx, dcolumn, amsmath, amsthm, amssymb, epstopdf, float}
\begin{document}
\title{Ultrafast X-Ray Induced Changes of the Electronic and Magnetic Response of Solids Due to Valence Electron Redistribution}
\author{Daniel J. Higley}
\email{dhigley@slac.stanford.edu}
\affiliation{SLAC National Accelerator Laboratory, 2575 Sand Hill Road, Menlo
Park, California 94025, USA}
\affiliation{Department of Applied Physics, Stanford University,
Stanford, California 94305, USA}
\author{Alex H. Reid}
\affiliation{SLAC National Accelerator Laboratory, 2575 Sand Hill Road, Menlo
  Park, California 94025, USA}
\author{Zhao Chen}
\affiliation{SLAC National Accelerator Laboratory, 2575 Sand Hill Road, Menlo
Park, California 94025, USA}
\affiliation{Department of Physics, Stanford University,
  Stanford, California 94305, USA}
\author{Lo{\"i}c Le Guyader}
\affiliation{SLAC National Accelerator Laboratory, 2575 Sand Hill Road, Menlo
  Park, California 94025, USA}
\affiliation{European X-Ray Free-Electron Laser Facility GmbH, Hozkoppel 4, 22869 Schenefeld, Germany}
\author{Olav Hellwig}
\affiliation{San Jose Research Center, HGST a Western Digital Company, 3403 Yerba Buena Rd., San Jose, California 95135, USA}
\affiliation{Institute of Physics, Chemnitz University of Technology, 09107 Chemnitz, Germany}
\affiliation{Institute of Ion Beam Physics and Materials Research, Helmholtz-Zentrum Dresden-Rossendorf, 01328 Dresden, Germany}
\author{Alberto A. Lutman}
\affiliation{SLAC National Accelerator Laboratory, 2575 Sand Hill Road, Menlo
  Park, California 94025, USA}
\author{Tianmin Liu}
\affiliation{SLAC National Accelerator Laboratory, 2575 Sand Hill Road, Menlo
Park, California 94025, USA}
\affiliation{Department of Physics, Stanford University,
  Stanford, California 94305, USA}
\author{Padraic Shafer}
\affiliation{Lawrence Berkeley National Laboratory, Berkeley, California
94720, USA}
\author{Tyler Chase}
\affiliation{SLAC National Accelerator Laboratory, 2575 Sand Hill Road, Menlo
Park, California 94025, USA}
\affiliation{Department of Applied Physics, Stanford University,
Stanford, California 94305, USA}
\author{Georgi L. Dakovski}
\affiliation{SLAC National Accelerator Laboratory, 2575 Sand Hill Road, Menlo
Park, California 94025, USA}
\author{Ankush Mitra}
\affiliation{SLAC National Accelerator Laboratory, 2575 Sand Hill Road, Menlo
Park, California 94025, USA}
\affiliation{Department of Physics, University of Warwick, Coventry CV4 7AL, United Kingdom}
\author{Edwin Yuan}
\affiliation{SLAC National Accelerator Laboratory, 2575 Sand Hill Road, Menlo
  Park, California 94025, USA}
\affiliation{Department of Applied Physics, Stanford University,
Stanford, California 94305, USA}
\author{Justine Schlappa}
\affiliation{European X-Ray Free-Electron Laser Facility GmbH, Hozkoppel 4, 22869 Schenefeld, Germany}
\author{Hermann A. D{\"u}rr}
\affiliation{SLAC National Accelerator Laboratory, 2575 Sand Hill Road, Menlo
  Park, California 94025, USA}
\author{William F. Schlotter}
\affiliation{SLAC National Accelerator Laboratory, 2575 Sand Hill Road, Menlo
Park, California 94025, USA}
\author{Joachim St{\"o}hr}
\email{stohr@slac.stanford.edu}
\affiliation{SLAC National Accelerator Laboratory, 2575 Sand Hill Road, Menlo
Park, California 94025, USA}
\begin{abstract}
  We report a novel mechanism, consisting of redistribution of valence electrons near the Fermi level, during interactions of intense femtosecond X-ray pulses with a Co/Pd multilayer. The changes in Co 3d valence shell occupation were directly revealed by fluence-dependent changes of the Co L$_3$ X-ray absorption and magnetic circular dichroism spectra near the excitation threshold. The valence shell redistribution arises from inelastic scattering of high energy Auger electrons and photoelectrons that lead to transient holes below and electrons above the Fermi level on the femtosecond time scale. The valence electron reshuffling effect scales with the energy deposited by X-rays and within 17 fs extends to valence states within 2 eV of the Fermi level. As a consequence the sample demagnetizes by more than twenty percent due to magnon generation.
\end{abstract}



\maketitle



Optical pump-probe techniques have revolutionized understanding of the dynamics of solid-state matter on picosecond and sub-picosecond timescales. Energy absorbed through valence electronic transitions can be observed to equilibrate and transfer to other degrees of freedom, such as phonons \cite{cho1990subpicosecond, chase2016ultrafast}, magnons \cite{kirilyuk2010ultrafast}, or as transitions of phase \cite{cavalleri2001femtosecond}. While optical techniques have provided a wealth of information about phenomena emerging 50 fs or longer after excitation, probing dynamics below this range has proven more challenging \cite{koopmans2000ultrafast, bigot2009coherent}. The initial driving mechanism of phenomena such as ultrafast demagnetization has remained unclear. 

X-ray free electron lasers have enabled the development of high pulse energy ultrashort X-ray pulses with durations as short as a few fs \cite{emma2010first, lutman2018high}. With soft X-ray excitation, energy is deposited into a material on a few femtosecond timescale without memory of the exciting pulse. This opens a new avenue to exploring material dynamics on ultrashort timescale using all X-ray techniques.

Ultrafast demagnetization is an excellent phenomenon to study with such all X-ray techniques. Despite intense study since its discovery more than twenty years ago \cite{beaurepaire1996ultrafast}, the mechanism(s) driving ultrafast demagnetization is still hotly debated \cite{koopmans2010explaining, battiato2010superdiffusive, shokeen2017spin, bonetti2016thz}. Recent experimental works have hinted at the importance of magnons \cite{turgut2016stoner, zusin2018direct, eich2017band} and understanding the first 30 fs after electronic excitation \cite{shokeen2017spin, tengdin2018critical, gort2018early}. Indirect evidence has also indicated significant energy transfer to the magnetic system within the first 30 fs \cite{tengdin2018critical}. However, experiments that can unambiguously probe magnetism on this $<$30 fs timescale are needed to determine whether energetic magnons drive ultrafast demagnetization.

X-ray magnetic circular dichroism spectroscopy (XMCD) is a well-established technique that gives element-specific information on the electronic and magnetic properties of unoccupied valence states \cite{stohr2006magnetism, de2008core}. Here, we perform X-ray fluence-dependent X-ray magnetic circular dichroism absorption spectroscopy (XMCD) of Co/Pd magnetic multilayers \cite{hellwig2007domain} across the Co L$_3$ resonance at 778 eV.  In our case, we use a single X-ray pulse which acts as both pump and probe of the sample dynamics. The averaged pump-probe delay is 15 fs, as determined by averaging the delay between pump and probe for all possible positive delays and weighting by the X-ray pump and probe strengths. We find that the $\approx$ 250 meV per atom deposited by the 39 fs Full Width at Half Maximum (FWHM) duration X-ray pulse is transferred to valence electrons within 2 eV of the Fermi level in 17 fs, and the sample demagnetizes by 22 percent averaged over the X-ray pulse duration. We thus probe, with the power of X-ray magnetic circular dichroism spectroscopy, the early stages of demagnetization. We note that the disappearance of resonant X-ray magnetic scattering with high X-ray fluence has been observed with similar samples, but with incident fluences higher than we investigate here \cite{muller2013breakdown, wu2016elimination}. For those higher fluences, effects such as stimulated emission and ionization-induced absorption edge shifts have been proposed to be important.



\section*{Results}


\begin{figure}
\centerline{\includegraphics[]{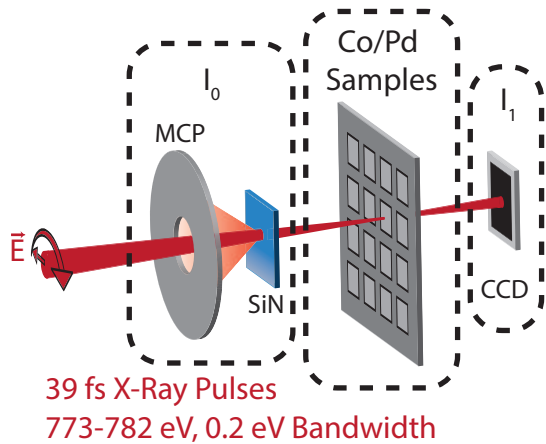}}
\caption{Experimental setup for high fluence X-ray absorption spectroscopy with circular polarization at LCLS. The Delta undulator produced circularly polarized X-rays which are then monochromatized. The total energy of these X-rays is detected with the fluorescence-based I$_0$ detector before they are focused on a Co/Pd sample. The X-ray pulse transmitted through the sample is attenuated before being detected with a CCD (I$_1$).}
\label{fig:setup}
\end{figure}

To measure the fluence dependence of X-ray absorption across the Co L$_3$ edge, we used the technique developed in \cite{higley2016femtosecond} with a few modifications to accommodate our use of high and variable incident X-ray fluence (see Methods). The key components of the experimental setup are shown schematically in Fig. \ref{fig:setup}. Circularly polarized X-ray pulses produced by the X-ray free electron laser were monochromatized to a bandwidth of $200$ meV. These X-ray pulses passed through a fluorescence-based relative X-ray pulse energy detector (I$_0$) \cite{higley2016femtosecond} before being focused onto Co/Pd magnetic multilayer samples. The samples had a metal layer sequence of Ta(1.5)Pd(3)[Co(0.6)Pd(0.6)]x38Pd(2), where the thicknesses in parentheses are in nm, and were magnetized out-of-plane. The total energy of X-ray pulses transmitted through the samples was detected with a CCD (I$_1$). The absolute X-ray pulse energy at the sample was determined using the CCD and a gas-based detector upstream of the samples (see Methods).

To calculate the X-ray excitation fluence, we have to take into account that both the excitation profile and probing profile are created by the same X-ray spot on the sample. The X-ray spot is approximately Gaussian and the X-ray fluence varies over this Gaussian spot. We quote the X-ray fluence as the fluence averaged over the X-ray pulse profile and weighted by the X-ray fluence at each point on the sample. This averaged X-ray fluence is one half of the peak X-ray fluence



\begin{figure}[htb]
\centerline{\includegraphics[]{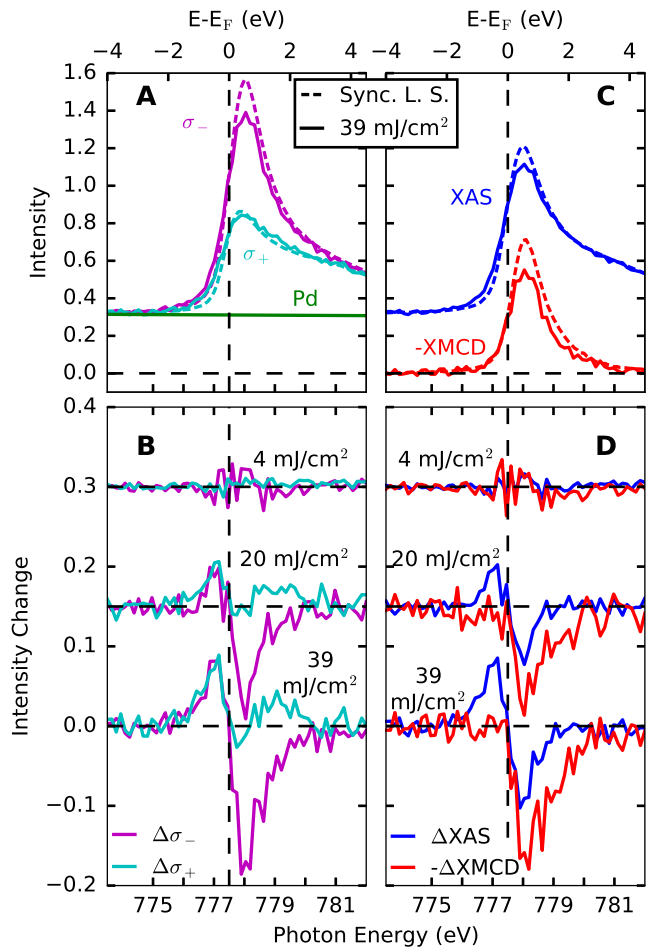}}
\caption{Fluence dependence of X-ray absorption around the Co L$_3$ edge of Co/Pd. Absorption spectra were recorded in the low fluence limit at the ALS synchrotron light source (labeled ``Sync. L. S.''), and with variable incident X-ray fluence at LCLS. The location of the Fermi level is also indicated with dashed vertical lines. (A) X-ray absorption spectra recorded with parallel ($\sigma_-$) and anti-parallel ($\sigma_+$) orientations of X-ray polarization and sample magnetization. Also shown is the contribution of the Pd absorbance to the total sample absorbance. (B) Difference of X-ray absorption spectra recorded at varying incident X-ray fluence at LCLS relative to those recorded in the low fluence limit. (C) XAS and XMCD spectra calculated from the data shown in (A). (D) Difference of XAS and XMCD recorded with varying incident X-ray fluence at LCLS relative to the low fluence limit.}
\label{fig:overall}
\end{figure}

Fig. \ref{fig:overall} shows the dependence of the absorption spectra around the Co L$_3$ resonance on incident X-ray fluence. At this resonance, X-ray absorption excites Co 2p$_{3/2}$ core electrons into unoccupied valence states which are primarily of 3d character. Spectra were recorded in the low fluence limit at the ALS synchrotron light source \cite{young2001first} (dashed spectra labeled ``Sync. L. S.''), and compared to those recorded with variable incident X-ray fluence at LCLS. The position of the Fermi level is determined as the zero crossing of the change in XAS, between where the change in XAS is positive below the absorption resonance, and negative at the peak of the absorption resonance, as described further below. The position of the Fermi level is indicated in Fig. \ref{fig:overall} with dashed vertical lines at 777.5 eV. The incident X-ray photon energy is shown on the bottom x-axes of Fig. \ref{fig:overall}, while the distance from the Fermi level is shown on the top x-axes. Above the Fermi level, the incident X-rays have sufficient photon energy to excite Co 2p$_{3/2}$ core electrons into unoccupied valence states.

Fig. \ref{fig:overall}a shows, for the low fluence limit (Sync. L. S.) and 39 mJ/cm$^2$ incident fluence cases, the dependence of X-ray absorption intensity on photon energy and the relative orientation of X-ray polarization and sample magnetization. One case of relative orientation of sample magnetization and X-ray polarization ($\sigma_-$) gives a large absorption resonance due to efficient excitation into the unoccupied states with the dominant spin orientation, while the other ($\sigma_+$) gives a much weaker absorption resonance. Also shown is the strength of the absorption due to the Pd in the sample (as calculated using \cite{henke1993x}). Fig. \ref{fig:overall}b shows the difference between the X-ray absorption measured at LCLS and that measured at the synchrotron light source as a function of the incident fluence at LCLS. For an incident X-ray fluence of 4 mJ/cm$^2$, there are no spectral changes outside of the noise of the measurement. For an incident X-ray fluence of 39 mJ/cm$^2$, X-ray absorption changes of more than ten percent of the resonant X-ray absorption magnitude occur.

Usually, spectra such as those shown in Fig. \ref{fig:overall}a are processed into and shown as so-called XAS and XMCD spectra. The XAS spectrum is the average of the $\sigma_+$ and $\sigma_-$ spectra, while the XMCD spectrum is their difference. In the low fluence limit, the XAS spectrum reflects the sample's electronic structure, while the XMCD spectrum reflects its magnetic structure. The division of the lowest and highest fluence spectra into these components is shown in Fig. \ref{fig:overall}c. These spectra are calculated using the same data as that of Fig. \ref{fig:overall}a, and are simply a different view of that data. Fig \ref{fig:overall}d shows the difference between the XAS and XMCD recorded at LCLS and that recorded at the synchrotron light source as a function of the incident X-ray fluence at LCLS.

There are large fluence-dependent changes of the XAS and XMCD shown in Fig. \ref{fig:overall}c and d. In the XAS, the absorption increases with increasing incident X-ray fluence below the Fermi level (\textless 777.5 eV), while it decreases with increasing incident X-ray fluence above the Fermi level (\textgreater 777.5 eV). The XAS changes are localized to within 1.5 eV of the Fermi level and do not extend over the entire absorption resonance.

These fluence-dependent changes in XAS indicate shuffling of valence electrons from below to above the Fermi level. This results in an increased XAS below the Fermi level due to an increase in unoccupied states to excite into with an energy below the Fermi level. Similarly, the decrease in XAS above the Fermi level results from a decrease in the available unoccupied states to excite into above the Fermi level (increase in occupation of states above the Fermi level).

This valence electron shuffling process occurs when the high energy Auger electrons and photoelectrons, which are generated following X-ray absorption, inelastically scatter with the valence electrons in the material. This process can create hundreds of electron-hole pairs per absorbed X-ray photon as the several hundred eV Auger- and photo-electrons decay to near the Fermi level in an electron cascade. Signs of this process have been observed in X-ray emission \cite{vinko2010electronic, medvedev2011short, schreck2014reabsorption, beye2013stimulated} and X-ray pump--optical probe measurements \cite{riedel2013single}, as commonly used for diagnosing relative arrival times of optical and X-ray pulses in experiments with XFELs \cite{beye2012x}. For the $<$ 50 fs timescale of this experiment, the deposited electronic energy is primarily also stored as electronic energy \cite{koopmans2010explaining, tengdin2018critical}. Some electrons escape the material in this cascade, but the energy they carry away from the sample is much less than the deposited X-ray energy \cite{henke19770, day1981photoelectric}.

The XMCD shown in Fig. \ref{fig:overall}c and d decreases with increasing incident X-ray fluence at a faster rate than the XAS changes. In contrast to the XAS changes, the XMCD does not change significantly below the Fermi level. The decrease in XMCD extends to higher photon energies than the decrease in XAS, as shown in Fig. \ref{fig:overall}c and d. The degree of XMCD decrease is close to the degree of demagnetization following strong optical excitation of similar materials in a $\approx$ 15 fs timescale \cite{kuiper2014spin}. This indicates that the XMCD changes in this fluence range can be explained through X-ray-induced demagnetization. The high energy electrons created following X-ray absorption do not scatter in a spin dependent manner \cite{lassailly1994spin}, but lower energy secondary electrons can drive an X-ray-induced demagnetization in the same manner as with optical excitation.



This discussion shows that the fluence-dependent changes in resonant X-ray absorption can be explained by X-ray-induced valence electron redistribution and demagnetization within the X-ray pulse duration. A difficulty in quantitative understanding of the spectral changes arises, however, due to the strongly varying strength of X-ray absorption in the vicinity of the Co L$_3$ resonance with respect to photon energy and relative orientation of sample magnetization and X-ray polarization (as seen in Fig. \ref{fig:overall}a). Due to this variation in X-ray absorption strength, the degree of sample excitation, varies greatly with respect to these parameters for a given incident fluence. To make the data more straightforwardly interpretable, we also calculate spectra which are chosen to have an approximately constant absorbed X-ray fluence rather than incident X-ray fluence.

\begin{figure}
\centerline{\includegraphics[]{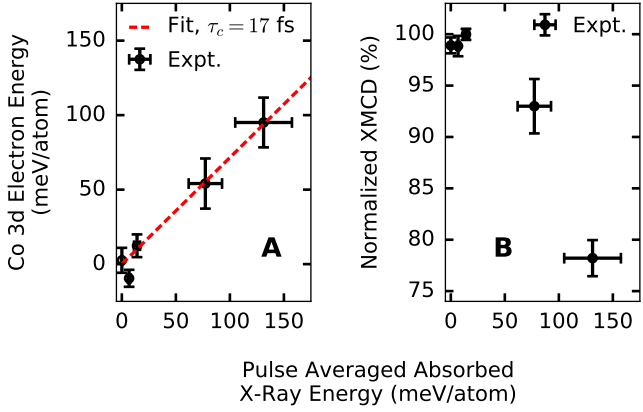}}
\caption{Quantification of the fluence-dependent X-ray absorption changes with respect to the pulse averaged energy absorbed in the Co/Pd multilayer (one half of the total absorbed energy). Vertical errorbars correspond to the standard error of the displayed quantities while horizontal errorbars correspond to the estimated 20 percent uncertainty in the absolute fluence calibration. (A) Co 3d electron energy averaged over the X-ray pulse as calculated from the fluence-dependent XAS changes. (B) Normalized XMCD strength averaged over its FWHM extent in photon energy.}
\label{fig:quant}
\end{figure}

\begin{figure}
\centerline{\includegraphics[]{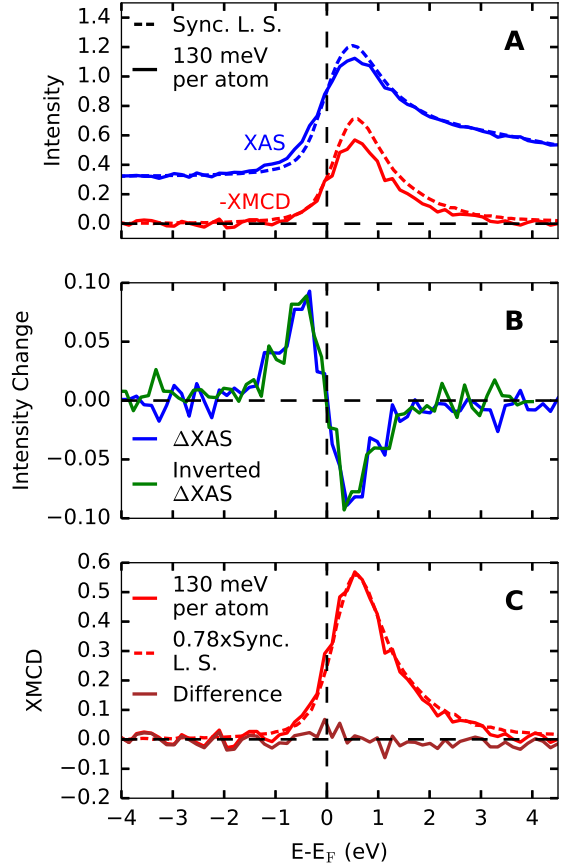}}
\caption{Comparison of spectra recorded with 130 meV/atom pulse averaged X-ray energy absorbed in the Co/Pd multilayer (260 meV/atom total absorbed energy, 15 mJ/cm$^2$ total absorbed fluence) to those recorded on the same sample in the low fluence limit at a synchrotron light source. (A) XAS and XMCD. (B) Difference of XAS relative to the low fluence limit. Also shown is this XAS difference inverted about the origin. (C) Difference of XMCD relative to the low fluence limit. Also shown is the original XMCD scaled by a factor.}
\label{fig:constant}
\end{figure}

Fig. \ref{fig:quant} shows a quantification of the absorption changes in these spectra recorded with a constant absorbed X-ray energy density (absorbed X-ray fluences were converted to absorbed X-ray energies per atom using the atomic densities given in \cite{haynes2014crc}). The spectral changes represent the state of the sample averaged over the X-ray pulse duration. The deposited X-ray energy averaged over the X-ray pulse duration is one half of the total deposited X-ray energy. We therefore quantify the spectral changes with respect to the pulse averaged absorbed X-ray energy, as shown in Fig. \ref{fig:quant}. This is in contrast to the spectra shown in Fig. \ref{fig:overall}, which were given with respect to the total incident X-ray fluence. The vertical errorbars of Fig. \ref{fig:quant} are standard errors of the displayed quantities. The horizontal errorbars correspond to the estimated 20 percent uncertainty in the calibration of the absolute X-ray fluence (the relative uncertainty for different X-ray fluence points is much less). The data extends up to a pulse averaged absorbed X-ray energy of 130 meV/atom. We note that the threshold dose for melting of these Co/Pd multilayers is 0.615 eV/atom (see methods). For solids, permanent sample damage is usually observed for samples exposed to single-shot X-ray doses larger than the melting threshold \cite{hau2007damage, hau2010interaction}.

Fig. \ref{fig:quant}a shows an estimation of the valence excitation strength as a function of the pulse averaged deposited X-ray energy using the XAS spectra and the XAS sum rule. This rule states that 
$N_h = k[L_3+L_2]$,
where $N_h$ is the number of Co 3d holes per atom, $L_3$ and $L_2$ are the integrals over the resonant components of the Co L$_3$ and L$_2$ edges, and $k$ is a proportionality constant \cite{stohr2006magnetism}. As we measured XAS changes only at the Co L$_3$ resonance, we approximated this sum rule as
$N_h = mL_3$,
where $m$ is another proportionality constant. To calculate $m$, we took $N_h$ to be 2.49 and subtracted a non-resonant background from the low fluence X-ray absorption spectra, as in \cite{chen1995experimental}. The change in number of Co 3d holes per unit energy is approximately given by an energy dependent version of this sum rule,
$\Delta N_h(E) = m\Delta L_3(E)$,
where $E$ is the particular energy of interest, and $\Delta L_3(E)$ is the change in XAS at that energy. The change in Co 3d electron energy per atom, $\Delta V$, is then given by multiplying by the distance below the Fermi level and integrating over energy:
\begin{equation}\label{eq:elec_energy}
\Delta V = -m\int (E-E_f) \Delta L_3(E)dE.
\end{equation}
Performing this integration from 2 eV below to 2 eV above the Fermi level, we obtain the results shown in Fig. \ref{fig:quant}a. The data is well represented by a linear fit with the spectrally calculated valence electron energy being 72 percent of the total deposited X-ray energy. This represents the degree to which the deposited X-ray energy has been transferred to electrons within 2 eV of the Fermi level within the X-ray pulse duration.

We further used the spectrally detected valence electron energy as a function of pulse averaged absorbed X-ray energy to estimate the duration of the cascade from initial high energy X-ray excitation to valence excitations within 2 eV of the Fermi level. To do so, we calculated, within a simple model and for the parameters of this experiment, the cascade duration, $\tau_c$, which gives a spectrally detected electron energy within 2 eV of the Fermi level which is 72 percent of the pulse averaged absorbed X-ray energy, as for the fit to the data of Fig. \ref{fig:quant}a. In this simple model, the electronic energy within 2 eV of the Fermi level was given by $v(t) = (I*c)(t)$, where $*$ denotes convolution, $c(t)$ gives the fraction of energy transferred to within 2 eV of the Fermi level following instantaneous X-ray excitation at $t = 0$, and $I(t)$ is the intensity of the X-ray pulse at the sample. $c(t)$ was modeled as a linear increase from zero at $t<0$ to one at $t>\tau_c$, with $\tau_c$ being the duration of the electron cascade.  $I(t)$ was modeled as the convolution of a flattop function with 25 fs FWHM (representing the pulse produced by LCLS) and a Gaussian with 34 fs FWHM (representing the pulse broadening due to the monochromator). We note that this neglects the spiky temporal structure of the XFEL pulses which were produced through self-amplified spontaneous emission \cite{saldin2010statistical}. The spectrally detected valence electron energy, $s$, was then given by averaging the valence electron energy over the X-ray pulse in time, $s = (\int dt I(t)v(t))/(\int dt I(t))$. Within this model, a cascade duration of $\tau_c = 17$ fs gives a spectrally detected electron energy within 2 eV of the Fermi level which is 72 percent of the pulse averaged absorbed X-ray energy, as in the fit to the data of Fig. \ref{fig:quant}a. This extracted X-ray-induced electron cascade duration of 17 fs is in reasonable agreement with the predicted duration of transfer of electronic energy from initial several hundred eV excitations to excitations with energies of 10 eV or less following soft X-ray excitation of condensed matter \cite{ziaja2005unified, medvedev2015femtosecond}.

Fig. \ref{fig:quant}b shows the XMCD as a function of the absorbed X-ray energy per atom. The strength of the XMCD decreases with increasing absorbed X-ray energy. For the highest fluence case of 130 meV per atom pulse averaged absorbed X-ray energy, the XMCD has decreased in magnitude by 22 percent.


Fig. \ref{fig:constant} shows the observed spectral changes with 130 meV/atom pulse averaged absorbed energy in the Co/Pd multilayer. Unlike in the case with constant incident fluence, the XAS changes observed below and above the Fermi level are now equal and opposite in shape and magnitude. This is further illustrated by plotting the XAS changes inverted about the Fermi level ($-\Delta XAS(-E)$), as shown in Fig. \ref{fig:constant}b. The inverted XAS changes overlap the measured XAS changes well. The similarity of these XAS changes below and above the Fermi level further reinforces their shared origin of shuffling of valence electrons from below to above the Fermi level. Further, the equal strengths of the increase in XAS below the Fermi level and decrease in XAS above the Fermi level show that the total number of 3d holes in Co is conserved during the X-ray pulse, and that there are no great changes in the electronic density of states within the X-ray pulse duration, as may be expected, for example, due to a change in the exchange splitting.

The XMCD in Fig. \ref{fig:constant}c is reduced in strength relative to its low X-ray fluence limit. Comparing this to a scaled version of the XMCD measured in the low fluence limit reveals no XMCD line shape changes outside of the experimental error except, possibly, a small deviation around the Fermi level. An important observation is that the XMCD changes extend through energies well above the Fermi level, even where the XAS is negligibly changed. This corresponds to the increase in $\sigma_+$ absorption intensity and equal magnitude decrease in $\sigma_-$ absorption intensity at 1.5 eV above the Fermi level in Fig. \ref{fig:overall}b. These changes cannot be explained through increases in the occupations of states with a specific spin polarization above the Fermi level, an effect which only decreases X-ray absorption intensity. Instead, the spin polarization of the density of states has decreased. Such a change can occur through significant magnon generation, as recently observed in time- and spin-resolved photoemission \cite{eich2017band}, and in temperature-dependent measurements of magnetic structure \cite{schneider1991spin, kachel2009transient}. That the XMCD of Fig. \ref{fig:constant}c is nearly unchanged in shape relative to the low fluence limit shows that this ultrafast magnon generation is the dominant contribution to the demagnetization observed in this experiment. Our results show that this magnon generation process is remarkably fast, already being the dominant contributor to demagnetization within 15 fs of X-ray excitation.


\section*{Discussion}
In summary, we have presented X-ray fluence-dependent changes of X-ray absorption at the Co L$_3$ edge of Co/Pd magnetic multilayers. The spectral changes reflect a transfer of deposited X-ray energy to valence electrons within 2 eV of the Fermi level in 17 fs and demagnetization of more than twenty percent due to magnon generation. Our study shows that X-ray-induced dynamics can be used to gain new insights into ultrafast processes, such as ultrafast demagnetization. In other cases, it is desirable to use high intensity X-ray pulses to probe sample characteristics without modification of the sample properties through X-ray excitation \cite{bostedt2016linac, wang2012femtosecond, kroll2018stimulated}. Our study sets limits on when this is possible.

\section*{Methods}

\subsection{Co/Pd Samples}
The Co/Pd samples were sputter deposited onto 100 nm thick Si$_3$N$_4$ membranes. They had a metal layer sequence of Ta(1.5)Pd(3)[Co(0.6)Pd(0.6)]x38Pd(2), where the thicknesses in parentheses are in nm. During measurement, the samples were magnetized out-of-plane with an applied magnetic field of 350 mT. This field was sufficient to saturate the magnetization out-of-plane, as verified through hysteresis loop measurements.

\subsection{High Fluence X-Ray Magnetic Circular Dichroism Absorption Spectroscopy}

We performed this experiment at the SXR hutch \cite{schlotter2012soft} of the LCLS X-ray free electron laser \cite{emma2010first, bostedt2016linac}.  The Delta undulator was operated in the diverted beam scheme, providing circularly polarized X-ray pulses with $200$ $\mu$J pulse energy and 25 fs FWHM pulse duration \cite{lutman2016polarization}. A grating monochromator then filtered these X-rays to a bandwidth of $200$ meV and broadened the pulses by 34 fs FWHM \cite{heimann2011linac, heimann2018private}, resulting in a pulse length of 39 fs FWHM (as estimated numerically). After monochromatization, the gas monitor detector measured the absolute X-ray pulse energy \cite{tiedtke2014absolute}. A fluorescence-based X-ray detector then measured the shot-to-shot incoming pulse energy \cite{higley2016femtosecond}, before the X-rays traversed a variable attenuation solid attenuator system, which enabled adjustment of the X-ray fluence at the sample. Next, a pair of Kirkpatrick-Baez mirrors focused the X-rays onto the sample at normal incidence with a spot size of 11 by 21 $\mu$m FWHM, as measured by pinhole scans.  The X-rays transmitted through the sample were attenuated before being detected with a CCD (Andor DO940P-BN-T2). The CCD was operated in ``spectroscopy'' mode, where the signal is integrated over each pixel column of the CCD before read out. In this mode, the CCD could be read out at the full 120 Hz repetition rate of LCLS. When data was recorded with fluences near and above the threshold for permanent sample damage, samples were replaced at a rate of 2 Hz. In calculating spectra, we only used data recorded on samples which had not been exposed to fluences above a threshold well below that required for permanent X-ray absorption changes.

\subsection{Determination of Absolute Pulse Energy}
The absolute X-ray pulse energy at the sample was determined using the gas monitor detector upstream of the sample and the CCD downstream of the sample. Despite the fundamental physical differences in the operation of these detectors, they gave values for the absolute pulse energy at the sample which were within 20 percent of each other. In calculating the pulse energy at the sample, the efficiency or transmission of components between the pulse energy detector and the sample was taken into account. The transmission of attenuators inserted before or after the sample was measured. The efficiency of each of the soft X-ray mirrors was estimated to be 0.73, and the transmission of the SiN membrane used in the I$_0$ detector was calculated to be 0.59 for the measured photon energies \cite{henke1993x}.

\subsection{Melting Threshold of Co/Pd}
The threshold dose for melting of the Co/Pd multilayer samples, $D_{melt}$, was approximated as
\begin{equation}
D_{melt} = f_{Co}\left[H_{f,Co}+H_{Co}(T_{melt})-H_{Co}(T_0)\right]+f_{Pd}\left[H_{Pd}(T_{melt})-H_{Pd}(T_0)\right],
\end{equation}
where $f_x$ is the fraction of atomic species $x$ in the Co/Pd multilayers, $H_{f, Co}$ is the enthalpy of fusion of Co, $H_x(T)$ is the enthalpy of $x$ at temperature $T$, and $T_{melt}$ is the melting temperature of Co, which has a lower melting temperature than Pd \cite{haynes2014crc}. Using the melting temperatures and atomic densities given in \cite{haynes2014crc}, as well as the enthalpies given in \cite{chase1998nist, arblaster2018re}, we obtain $D_{melt} = $ 0.615 eV/atom. The enthalpies tabulated in \cite{arblaster2018re} have been interpolated to $T_{melt}$ where necessary.

\section*{Acknowledgments}
We acknowledge M. Beye, C. D. Pemmaraju, D. A. Reis, and P. Heimann for valuable discussions. We thank J. Aldrich for technical assistance. Use of the Linac Coherent Light Source, SLAC National Accelerator Laboratory, is supported by the U.S. Department of Energy, Office of Science, Office of Basic Energy Sciences under Contract No. DE-AC02-76SF00515. The Advanced Light Source is supported by the Director, Office of Science, Office of Basic Energy Sciences, of the U.S. Department of Energy under Contract No. DE-AC02-05CH11231.

\section*{Contributions}
D. J. H., A. H. R., Z. C., L. L. G., A. A. L., T. L., P. S., T. C., G. L. D., A. M., E. Y., J. S., H. A. D., W. F. S., and J. S. performed experiments. O. H. and Z. C. prepared samples. D. J. H. analyzed experimental data and performed calculations. D. J. H., A. H. R., and J. S. wrote the manuscript with input from all authors.

\bibliographystyle{naturemag}
\bibliography{nonlinear_xmcd_refs}

\end{document}